\documentclass[twocolumn,citeautoscript,prl,superscriptaddress,amsmath,amssymb,8pt]{revtex4-1}

\usepackage{graphicx}
\usepackage{color}
\usepackage[dvipsnames]{xcolor}
\usepackage{upgreek}
\usepackage{float}

\usepackage{upgreek}
\usepackage{amsmath,amssymb}
\newcommand{\C}[1]{\ensuremath{^{#1}}C}
\newcommand{\ket}[1]{\ensuremath{\left| #1 \right\rangle}}

\begin{document}

\title{Quantum probe hyperpolarisation of molecular nuclear spins}

\author{David A. Broadway}
\email{broadway@student.unimelb.edu.au}
\affiliation{Centre for Quantum Computation and Communication Technology, School of Physics, University of Melbourne, Parkville, VIC 3010, Australia}
\affiliation{School of Physics, University of Melbourne, Parkville, VIC 3010, Australia}	

\author{Jean-Philippe Tetienne}
\affiliation{Centre for Quantum Computation and Communication Technology, School of Physics, University of Melbourne, Parkville, VIC 3010, Australia}
\affiliation{School of Physics, University of Melbourne, Parkville, VIC 3010, Australia}	

\author{Alastair Stacey}
\affiliation{Centre for Quantum Computation and Communication Technology, School of Physics, University of Melbourne, Parkville, VIC 3010, Australia}
\affiliation{Melbourne Centre for Nanofabrication, Clayton, VIC 3168, Australia}

\author{James D. A. Wood}
\affiliation{Centre for Quantum Computation and Communication Technology, School of Physics, University of Melbourne, Parkville, VIC 3010, Australia}
\affiliation{School of Physics, University of Melbourne, Parkville, VIC 3010, Australia}
\affiliation{Department of Physics, University of Basel, Klingelbergstrasse 82 , 4056, Basel, Switzerland}

\author{David A. Simpson}
\affiliation{School of Physics, University of Melbourne, Parkville, VIC 3010, Australia}	

\author{Liam T. Hall}
\email{liam.hall@unimelb.edu.au}
\affiliation{School of Physics, University of Melbourne, Parkville, VIC 3010, Australia}

\author{Lloyd C. L. Hollenberg}
\email{lloydch@unimelb.edu.au}
\affiliation{Centre for Quantum Computation and Communication Technology, School of Physics, University of Melbourne, Parkville, VIC 3010, Australia}
\affiliation{School of Physics, University of Melbourne, Parkville, VIC 3010, Australia}

\begin{abstract}
	The hyperpolarisation of nuclear spins within target molecules is a critical and complex challenge in magnetic resonance imaging (MRI)\,\cite{ChristopherdeCharms2008} and nuclear magnetic resonance (NMR) spectroscopy\,\cite{Rossini2013}. Hyperpolarisation offers enormous gains in signal and spatial resolution which may ultimately lead to the development of molecular MRI and NMR \cite{Nikolaou2015}. At present, techniques used to polarise nuclear spins generally require low temperatures and/or high magnetic fields, radiofrequency control fields, or the introduction of catalysts or free-radical mediators\,\cite{Bajaj2003,Viteau2008,London2013,Bhattacharya2011}. The emergence of room temperature solid-state spin qubits has opened exciting new pathways to circumvent these requirements to achieve direct nuclear spin hyperpolarisation using quantum control\,\cite{London2013,Alvarez2015}. Employing a novel cross-relaxation induced polarisation (CRIP) protocol, we demonstrate the first external nuclear spin hyperpolarisation achieved by a quantum probe, in this case of $^1$H molecular spins in poly(methyl methacrylate). In doing so, we show that a single qubit is capable of increasing the thermal polarisation of $\sim 10^6$ nuclear spins by six orders of magnitude, equivalent to an applied magnetic field of $10^5$\,T. The technique can also be tuned to multiple spin species, which we demonstrate using both \C{13} and $^1$H nuclear spin ensembles. Our results are analysed and interpreted via a detailed theoretical treatment, which is also used to describe how the system can be scaled up to a universal quantum hyperpolarisation platform for the production of macroscopic quantities of contrast agents at high polarisation levels for clinical applications. These results represent a new paradigm for nuclear spin hyperpolarisation for molecular imaging and spectroscopy, and beyond into areas such as materials science and quantum information processing. 
\end{abstract}

\maketitle

With the prospect of molecular MRI\,\cite{Nikolaou2015} revolutionising many areas of research and clinical applications, the generation of hyperpolarised molecular targets attracts intense interest. Well known techniques include high-field/low temperature brute-force methods \cite{Bajaj2003}, dynamic nuclear polarisation \cite{London2013}, optical pumping \cite{Viteau2008}, and parahydrogen induced polarisation \cite{Bhattacharya2011}. On the other hand, the rapid advances in semiconductor spin qubit technology for quantum computing \cite{Morello2010,Zwanenburg2013} and quantum sensing \cite{Doherty2013,Mamin2013} has opened up the exciting possibility of hyperpolarising nuclear spins at a fundamental level via quantum mechanical protocols \cite{Knowles2016,Belthangady2013,Wang2013,Fischer2013,King2015,Chen2015,Alvarez2015,Liu2014,Rej2015,London2013}. Despite early progress \cite{Abrams2014,Chen2016a}, the application of quantum probe technology to this problem still faces a number of significant challenges. To be of practical use, a quantum probe must be capable of polarising a relatively large number of remote nuclear spins external to the probe substrate, ideally under ambient conditions. We address these challenges using a quantum spin probe in diamond as an entropy pump, and demonstrate polarisation of external molecular spin ensembles over relatively large volumes at room temperature, with the prospect of scaling up to a universal hyperpolarisation platform suitable for clinical applications. In contrast to existing methods, our quantum polarisation approach is tunable to a range of nuclear species, operates at room temperature, and is inherently free of radiofrequency (RF) fields and extraneous chemistry. 

\begin{figure*}
	\includegraphics[width=\textwidth]{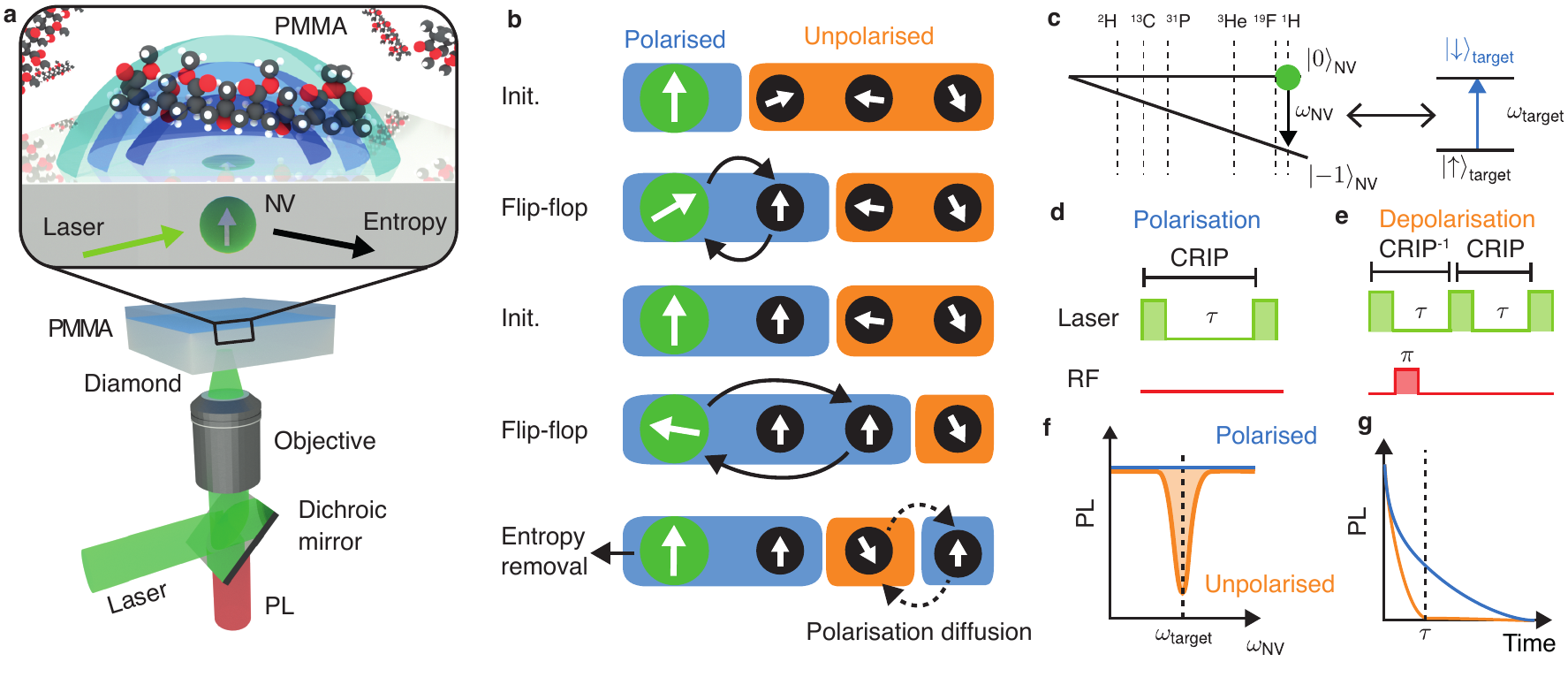}
	\caption{\textbf{Quantum probe hyperpolarisation of nuclear spin ensembles.}
		\textbf{a,} Schematic of the system showing a near-surface nitrogen-vacancy (NV) spin probe in diamond and a hydrogen nuclear spin target ensemble in molecular Poly(methyl methacrylate) (PMMA) on the surface. The NV probe is initialised by a green laser (532 nm), and read out via its photoluminescence (PL) signal. The shaded blue surfaces denote different regimes of polarisation capabilities arising from the spatial dependence of the nuclear spin coupling to the NV qubit.
		\textbf{b,} Schematic of cross relaxation induced polarisation (CRIP) implemented on a spin system illustrating the build up of polarisation from repeated application of the CRIP sequence. Diffusion effects act in competition with the CRIP entropy pumping mechanism, but also allow for polarisation at distances beyond that reachable via the hyperfine interaction.
		\textbf{c,} Energy-level diagram of the NV showing the relative positions of various target nuclear spin resonance conditions.
		\textbf{d,e,} The control sequences laser pulses in green, RF pulses in red) used for polarising a target spin ensemble using CRIP ({\bf d}) and for controlled depolarisation using the combined CRIP$^{-1} \times$ CRIP protocol ({\bf e}). \textbf{f,} Schematic showing the cross-relaxation spectrum obtained by measuring the PL during the CRIP (blue) or depolarisation (orange) sequence with a constant interaction time $\tau$, while scanning the NV frequency $\omega_{\rm NV}$.
		\textbf{g,} Similarly, the cross-relaxation curve is obtained by scanning $\tau$ with $\omega_{\rm NV}$ set at the resonance. \label{Fig: Intro}}
\end{figure*}

For this study, we employ a nitrogen-vacancy (NV) quantum spin probe in a diamond substrate (Fig.\,\ref{Fig: Intro}\textbf{a}) \cite{Doherty2013}. The protocol employs an external magnetic field, $B$, to tune the ground-state spin transition frequency of the NV ($\omega_{\rm NV}$) into resonance with target nuclear spins ($\omega_{\rm n}$) (Fig.\,\ref{Fig: Intro}\textbf{b}). For a given target species, the spin resonance condition is fulfilled at a magnetic field $B_*(\gamma_{\rm n})  \approx 2D / (\gamma_{\rm NV} - \gamma_{\rm n} )$ \cite{Broadway2016}, where $\gamma_\mathrm{n}$, $\gamma_\mathrm{NV}$ are the target and NV gyromagnetic ratios, and $D$ is the NV zero-field splitting (Fig.\,\ref{Fig: Intro}\textbf{c}). Entropy pumping is facilitated by repeated application of the cross relaxation induced polarisation (CRIP) sequence (Fig.\,\ref{Fig: Intro}\textbf{d}), wherein the NV spin is optically initialised into \ket{0} state and the NV-target hyperfine interaction is allowed to occur for a given period of time, $\tau$ (of order microseconds). The transfer of magnetisation caused by this interaction thus polarises the target spins into their \ket{\downarrow} state (Fig.\,\ref{Fig: Intro}\textbf{c}). For the sake of comparison, depolarisation may be facilitated by interleaving the initialisation of the NV spin into the opposite state \ket{-1} by the application of an RF $\pi$-pulse (Fig.\,\ref{Fig: Intro}\textbf{e}).

To quantitatively understand the effect of the CRIP protocol on the target spin ensemble, we developed a microscopic theory that explicitly includes the dipole interactions of ensemble spins and their interaction with a single NV quantum probe (see Supplementary Information for details). We define the polarisation of a spin at position $\mathbf{R}$ (relative to the NV) and time $t$ to be $P\left(\mathbf{R},t\right)$; with the evolution of $P(\mathbf{R},t)$ described by
\begin{eqnarray}
	\frac{\partial}{\partial t}P\,(\mathbf{R},t) &=&   \left(\beta\,\nabla^2 -u(\mathbf{R}) - \Gamma_{\text{SL}} \right) P(\mathbf{R},t)  + u(\mathbf{R}) , \,\,\,\,\,\,\,\,\,\label{cont}
\end{eqnarray} 
subject to an initial unpolarised state $P(\mathbf{R},t)=0$; where $u(\mathbf{R}) = A^2(\mathbf{R})/2\Gamma_2$ is the effective cooling coefficient resulting from the hyperfine coupling $A(\mathbf{R})$ with the NV spin, $\Gamma_2$ is the dephasing rate of the NV spin $\beta$ is the effective polarisation diffusion coefficient related to the intra-target interactions, and $\Gamma_\mathrm{SL}$ is the spin-lattice relaxation rate of the target spin ensemble. This formulation allows us to predict and describe the spatial extent of polarisation for a given target sample of arbitrary geometry. 

To probe the polarisation effect experimentally, we monitor the spin-dependent photoluminescence (PL) from the NV \cite{Hall2016,Wood2016,Wood2016a} during the laser pulses, which decays as a function of the CRIP sequence time as $e^{-\Gamma_{\rm tot}\tau}$. Here $\Gamma_{\rm tot}$ is the NV longitudinal relaxation rate, which can be expressed as the sum $\Gamma_\mathrm{tot} = \Gamma_\mathrm{bg} + \Gamma_\mathrm{CR}$, where $\Gamma_\mathrm{bg}$ is the background rate caused by lattice phonons or surface effects, and $\Gamma_\mathrm{CR}$ is due to cross-relaxation. The latter follows a Lorentzian dependence on the detuning between the probe and target transition frequencies\cite{Hall2016},
\begin{eqnarray}
	\label{eq:GammaCR}
	\Gamma_\mathrm{CR} &=& \frac{ A_P^2\Gamma_2}{ 2\Gamma_2^2+2\left(\omega_\mathrm{NV}-\omega_\mathrm{n}\right)^2},
\end{eqnarray}	
where $ A_P^2$ is the total hyperfine field variance seen by the NV due to the target ensemble, which is related to the polarisation distribution via
\begin{eqnarray}
	 A_P^2 &=& \frac{n_{\rm t}}{2} \int \left[1-P(\mathbf{R},t)\right]A^2(\mathbf{R})\,\mathrm{d}^3\mathbf{R},\,\,\,\,\,\,\,\,\label{pSol}
\end{eqnarray}
where $n_\mathrm{t}$ is the density of the target spin ensemble. The key indicator of significant polarisation is therefore a reduction in $\Gamma_\mathrm{CR}$, which manifests as the disappearance of the target ensemble's spectral feature from the cross-relaxation spectrum (Fig.\,\ref{Fig: Intro}\textbf{f}), and can be quantified by measuring the cross-relaxation curve at resonance (Fig.\,\ref{Fig: Intro}\textbf{g}).

\begin{figure}
	\includegraphics[width=\columnwidth]{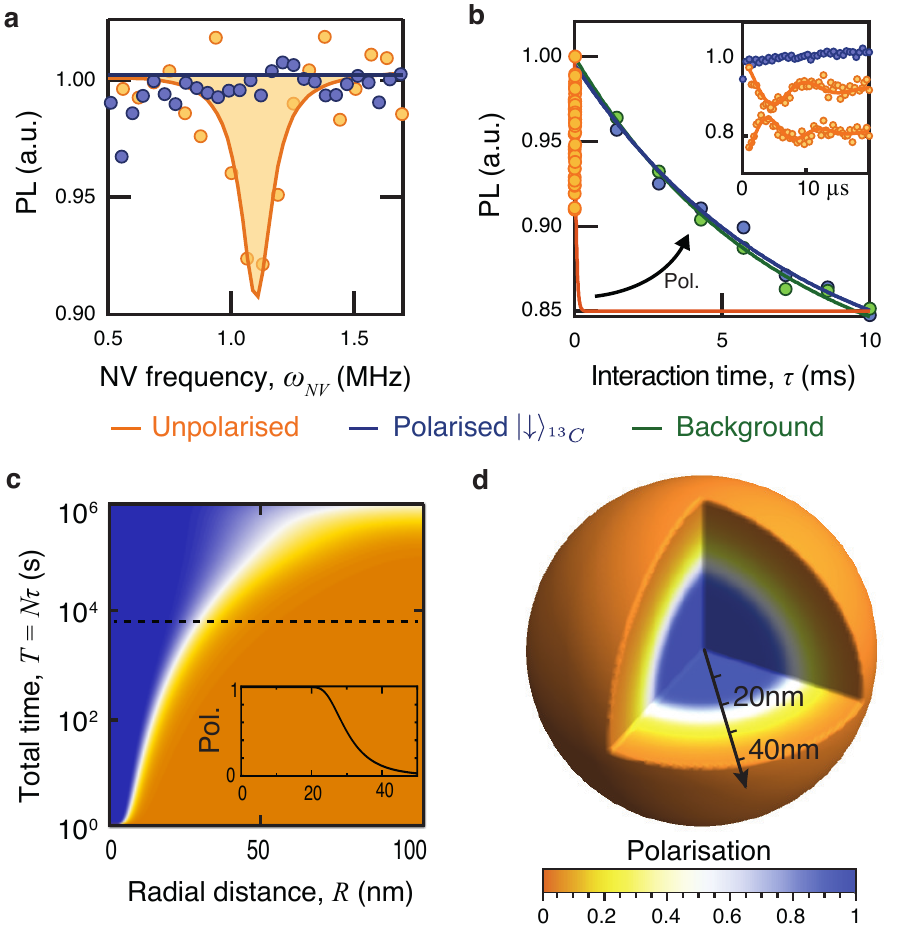}
	\caption{\textbf{Cross-relaxation induced polarisation of \C{13} spins in diamond.} 
	{\bf a,} Cross-relaxation spectra of a single NV spin near the $^{13}$C resonance ($\omega_{\rm NV}=1.1$~MHz), obtained with an interaction time $\tau=4~\mu$s using the CRIP sequence (blue) and the depolarisation sequence (orange, only the readout following the NV initialisation in $|0\rangle$ is shown). Sequences were repeated $N=10^5$ times at each point.
{\bf b,} Cross-relaxation curves obtained by increasing $\tau$ at the $^{13}$C resonance with the CRIP sequence (blue) and depolarisation sequence (orange), and off-resonance to obtain the background relaxation curve (green). Zoom-in at short times for the polarised (blue) and unpolarised case (orange, top and bottom curves correspond to the NV initialised in $|0\rangle$ and $|-1\rangle$, respectively).
{\bf c,} Calculated radial polarisation profiles relative to the NV spin (averaged over all angles), calculated from Eq. (1) for a random 1.1\% $^{13}$C spin ensemble for varying total polarisation times, $T=N\tau$. Inset: profile along dashed line, corresponding to $T=2$~h. 
{\bf d,} Three-dimensional representation of the polarisation distribution at $T=2$~h.
	} \label{Fig: Polarisation}
\end{figure}

Experimentally, we first demonstrate our technique on the inherent 1.1\% \C{13} spin ensemble surrounding a NV probe in the diamond substrate by tuning to the $^{13}$C resonant condition at $B_*(^{13}{\rm C}) = 1024.9$ G. Comparison of the cross-relaxation spectra for CRIP and depolarisation sequences (Fig.\,\ref{Fig: Polarisation}\textbf{a}) shows the complete removal of the \C{13} resonance peak for interaction times of $\tau = 4\,\upmu$s, indicating efficient polarisation of the nearest spins, as compared with the target prepared using the depolarising sequence. This is confirmed in the cross-relaxation curves as a function $\tau$ (Fig. 2b, inset), where the polarised case shows no evolution of the NV spin state, while the unpolarised case shows coherent flip-flops between the NV and the $^{13}$C spins.

To investigate the extent of the polarisation effect, we increase the interaction time $\tau$ so as to be sensitive to more remote $^{13}$C spins, up to the limit set by the NV centre's intrinsic spin-phonon relaxation rate, $\Gamma_{\rm bg}\approx200$~ms$^{-1}$. The resulting cross-relaxation curves obtained at the $^{13}$C resonance using the CRIP and depolarisation sequences are shown in Fig.\,\ref{Fig: Polarisation}\textbf{b}, from which we extract the total relaxation rate, $\Gamma_{\rm tot}$. By subtracting $\Gamma_{\rm bg}$ obtained from the off-resonance relaxation curve (Fig.\,\ref{Fig: Polarisation}\textbf{b}, green data), we deduce the $^{13}$C-induced relaxation rate $\Gamma_{\rm CR}=\Gamma_{\rm tot}-\Gamma_{\rm bg}$, which decreases from $\Gamma_{\rm CR}^{\rm unpol}\approx250$~ms$^{-1}$ with the depolarisation sequence, to below the noise floor of the measurement after 5 hours of CRIP, $\Gamma_{\rm CR}^{\rm pol}\lesssim19$~s$^{-1}$. We use Eq.\,\eqref{cont} (with $\beta = 0.0335$ nm$^2$s$^{-1}$ corresponding to the given \C{13} density) to calculate the time-dependence of the radial polarisation profile for total polarisation times of 1-$10^6$\,s, as depicted in Fig\,\ref{Fig: Polarisation}\textbf{c}. By relating the spatial polarisation distribution, $P(\mathbf{R},t)$, to the cross-relaxation rate, $\Gamma_\mathrm{CR}$, via Eq. (\ref{eq:GammaCR}), we find the theoretical results are consistent with the experiment for polarisation times in excess of two hours (Fig.\,\ref{Fig: Polarisation}\textbf{c}, dashed line). Examination of the spatial polarisation distribution (Fig.\,\ref{Fig: Polarisation}\textbf{c}, inset, and Fig.\,\ref{Fig: Polarisation}\textbf{d}) implies a polarisation level of more than 99\% within 21\,nm of the NV, equating to a $6\times 10^6$-fold  increase on thermal polarisation for $3\times 10^5$ spins.

\begin{figure}
	\includegraphics{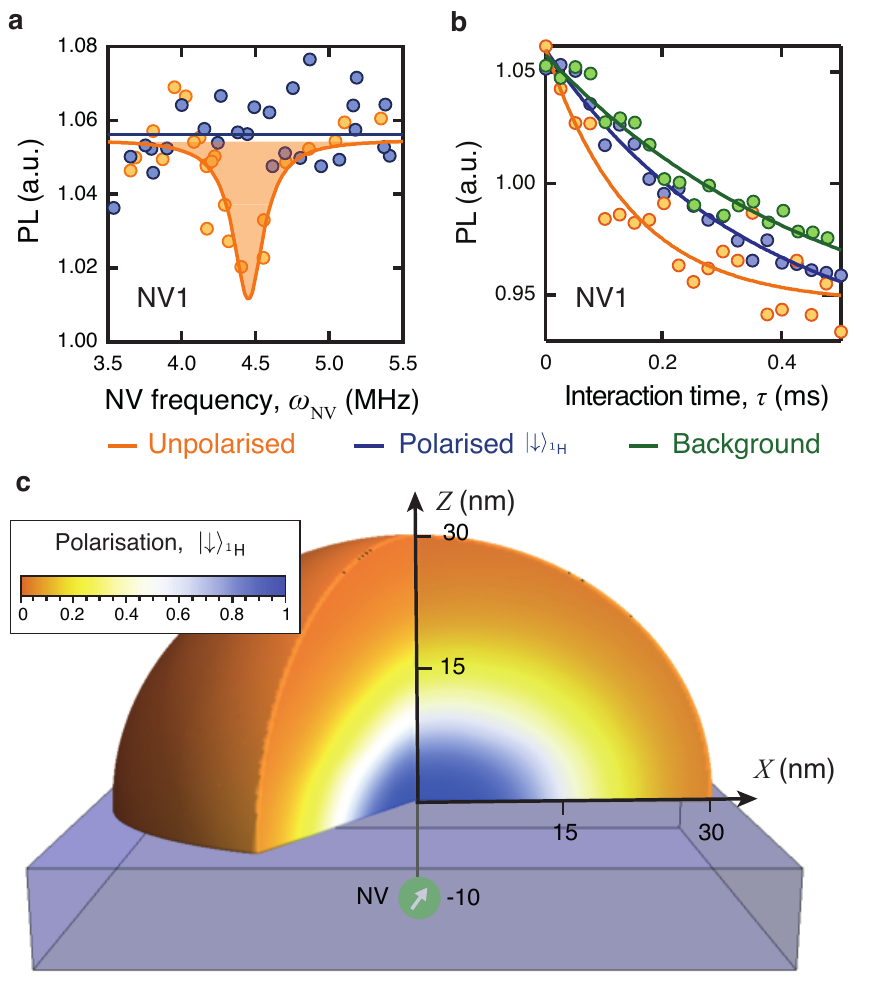}
	\caption{\textbf{Polarisation of external molecular $^1$H spins.} {\bf a,} Cross-relaxation spectra near the $^1$H resonance ($\omega_{\rm NV}=4.4$~MHz), for a single NV spin located 10 nm below a PMMA layer, obtained using $\tau=20~\mu$s and $N=10^5$ with the CRIP sequence (blue) and the depolarisation sequence (orange). 
{\bf b,} Cross-relaxation curves obtained by increasing $\tau$ at the $^1$H resonance with the CRIP sequence (blue) and depolarisation sequence (orange), and off-resonance to obtain the background relaxation rate (green).
{\bf c,} Three-dimensional representation of the $^1$H spin polarisation distribution in the PMMA, calculated from Eq. (\ref{cont}) in the steady state. \label{Fig: H pol}}
\end{figure}

With the basic protocol established, we now move to the polarisation of molecular $^1$H nuclear spins external to the diamond crystal. A solution of poly(methyl methacrylate), PMMA, was applied directly to a diamond substrate \cite{Mamin2013} with single NV spin probes located 8-12~nm below the surface. CRIP was applied with the external magnetic field tuned to resonance at $ B_* (^1{\rm H}) = 1026.2$ G. With a much higher diffusion constant and spin-lattice relaxation rate ($\beta=781~\mathrm{nm}^2s^{-1}$, $\Gamma_\mathrm{SL} = 1$s$^{-1}$) relative to the intrinsic $^{13}$C case, the $^1$H system effectively reaches steady-state within a few seconds. Application of the CRIP sequence to NV1 (data for other NVs are shown in the Supplementary Information) for $\tau=20\,\upmu$s  (Fig.\,\ref{Fig: H pol}\textbf{a}) shows the removal of the hydrogen spectral feature (blue), as compared with the depolarising sequence (orange). From the cross-relaxation curves after 1 hour of CRIP (Fig\,\ref{Fig: H pol}\textbf{b}), we extract $^1$H-induced rates for the unpolarised ($\Gamma_{\rm CR}^{\rm unpol}$) and polarised ($\Gamma_{\rm CR}^{\rm pol}$) PMMA $^1$H spin ensembles to be 2.71 ms$^{-1}$ and 0.96 ms$^{-1}$, respectively. The ratio $\Gamma_{\rm CR}^\mathrm{unpol}/\Gamma_{\rm CR}^\mathrm{pol} = 2.8(3)$ (consistent with the value of 2.4(3) obtained using another NV), is in good agreement with the solution to Eq.\,(\ref{cont}) for the PMMA ensemble, which gives a ratio of $2.2$ in the steady state. The corresponding spatial polarisation distribution is shown in Fig.\,\ref{Fig: H pol}\textbf{c}, indicating  that the system reaches 50\% average polarisation over a volume of $\sim$ (26 nm)$^3$. Thus, we conclude that the single spin quantum probe has increased the average polarisation of roughly a million hydrogen spins by	some six orders of magnitude over the room temperature Boltzmann thermal background.

\begin{figure}
	\centering
	\includegraphics{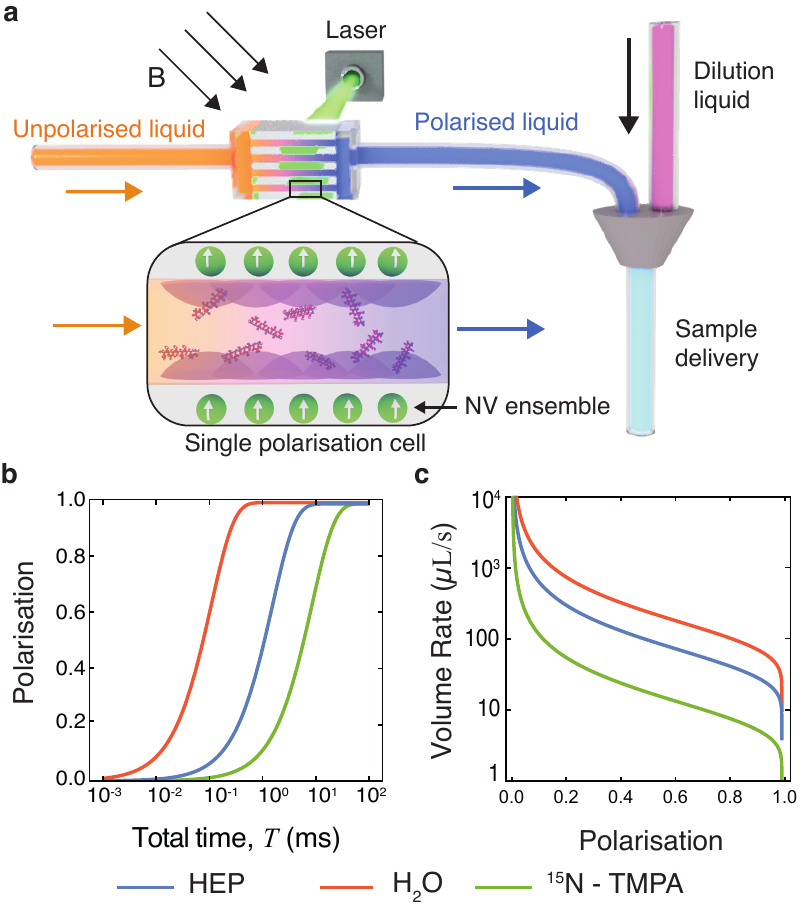}
	\caption{\textbf{Scale-up for a universal MRI contrast agent hyperpolarisation platform.} \textbf{a,} Schematic of a quantum polarisation stack comprising multiple diamond membranes, each containing NV array layers on both sides, in a homogeneous magnetic field tuned to the nuclear gyromagnetic ratio of the target agent spin species. The unpolarised agent in concentrated  solution (orange) flows into the stack channels, where the liquid is polarised through the application of CRIP (via a pulsed laser). The out-flowing polarised liquid (blue) is then diluted for use. Zoomed schematic shows a single polarisation cell comprising a channel formed by dual diamond membranes each with a near-surface NV layer. \textbf{b,} Average polarisation level from a single polarisation cell, for various targets (HEP, H$_2$O, and $^{15}$N-TMPA),  calculated for varying polarisation times assuming perfect mixing of a 1 M target agent solution with a cell height of 1 $\upmu$m. \textbf{c,} Outflow rate (after dilution to~1mM for application delivery) from 10 polarisation cells at different levels of polarisation.\label{Fig: MRI}}	
\end{figure}

There is scope for improvement on these proof-of-concept results: for example, engineering NV depths to 5\,nm would increase the rate of target spin polarisation by an order of magnitude, and improvements in the inherent NV dephasing rate $\Gamma_2$ (e.g. via improved surface properties) will allow for more precise tuning to different nuclear spin species. As the protocol is all optical, scaling up for high-volume production could be achieved by stacking multiple NV arrays (Fig.\,\ref{Fig: MRI}\textbf{a}) and/or increasing the effective interaction area via surface patterning \cite{Kehayias2017}. The results presented here indicate that the CRIP protocol could produce macroscopic quantities of MRI contrast agents with high polarisation levels. For example, we consider $^{13}$C isotopically enriched HEP $\left(\text{hydroxyethyl\,propionate,\,}\mathrm{^{13}C_5H_{10}O_3}\right)$, a well-known MRI contrast agent \cite{Schmidt2017}. Using a single hyperpolarisation cell comprised of two NV arrays in diamond membranes separated by 1 $\upmu$m (see zoomed schematic in Fig.\,\ref{Fig: MRI}\textbf{a}; we assume an NV density of 4$\times 10^{11}$ cm$^{-2}$ over a 4 mm $\times$ 4 mm diamond surface \cite{Simpson2017}), the rate of polarisation transfer to a concentrated 1M precursor HEP solution is 4 $\upmu$L/s at a polarisation level of 80\%. The polarisation levels for different contrast agents in 1M precursor solutions are plotted against polarisation time (assuming perfect mixing occurs over these timescales) in Fig.\,\ref{Fig: MRI}\textbf{b}.
In Fig.\,\ref{Fig: MRI}\textbf{c}, we plot the final delivery rate after dilution to 1~mM for a stack of 10 cells, showing that delivery rates of order 100 $\upmu$L/s for clinical applications \cite{Golman2006} are achievable.

In summary, we have experimentally demonstrated hyperpolarisation of molecular nuclear spins under ambient conditions by employing a quantum spin probe entropy pump. The technique works at low field, room temperature, requires no RF fields, and operates directly on the target molecules without the need for catalysts or free radicals. With high polarisation rates and tunability, there are excellent prospects for scale-up of the system to produce macroscopic quantities of a range of contrast agents at polarisation levels required for molecular MRI/NMR. The technique can be extended to other nuclear spin species and may also offer new pathways in quantum information for initialisation of quantum simulators, or increasing the fidelity of operations through spin-bath neutralisation.

\textbf{Author Contributions:} The CRIP protocol was conceived by LTH. Experiments were performed by DAB and J-PT, with input from AS, DAS, JDAW, LTH, and LCLH. LTH developed the theory, with input from LCLH. DAB, J-PT, AS, and DAS prepared the samples. LCLH, DAB and LTH wrote the manuscript with input from all authors. LCLH supervised the project.

\textbf{Acknowledgements:} This work was supported in part by the Australian Research Council (ARC) under the Centre of Excellence scheme (project No. CE110001027). This work was performed in part at the Melbourne Centre for Nanofabrication (MCN) in the Victorian Node of the Australian National Fabrication Facility (ANFF). L.C.L.H. acknowledges the support of an ARC Laureate Fellowship (project No. FL130100119). J.-P.T acknowledges support from the ARC through the Discovery Early Career Researcher Award scheme (DE170100129) and the University of Melbourne through an Establishment Grant and an Early Career Researcher Grant. D.A.B is supported by an Australian Government Research Training Program Scholarship.
	
\textbf{Competing Interests:} The authors declare that they have no competing financial interests.

\textbf{Correspondence:} Correspondence and requests for materials should be addressed to L.C.L.H. (email: lloydch@unimelb.edu.au), D.A.B (email: broadway@student.unimelb.edu.au), or L.T.H (email: liam.hall@unimelb.edu.au).


\end{document}